\documentclass[letterpaper,conference]{IEEEtran}

\usepackage[T1]{fontenc}
\usepackage[latin9]{inputenc}
\usepackage{array}
\usepackage{float}
\usepackage{amsbsy}
\usepackage{amstext}
\usepackage{amssymb}
\usepackage{epstopdf}
\usepackage{graphicx}
\usepackage{subfig}
\usepackage{algorithm}
\usepackage{algpseudocode}
\usepackage{cite}
\usepackage{amsmath}

\makeatletter

\providecommand{\tabularnewline}{\\}
\floatstyle{ruled}
\newfloat{algorithm}{tbp}{loa}
\providecommand{\algorithmname}{Algorithm}
\floatname{algorithm}{\protect\algorithmname}

\@ifundefined{showcaptionsetup}{}{%
 \PassOptionsToPackage{caption=false}{subfig}}
\makeatother

\begin{document}

\title{Toward a Robust Sparse Data Representation for Wireless Sensor Networks}
\author{\IEEEauthorblockN{Mohammad Abu Alsheikh\IEEEauthorrefmark{1}\IEEEauthorrefmark{2},
Shaowei Lin\IEEEauthorrefmark{2},
Hwee-Pink Tan\IEEEauthorrefmark{3},
and Dusit Niyato\IEEEauthorrefmark{1}}

\IEEEauthorblockA{\IEEEauthorrefmark{1}School of Computer Engineering, Nanyang Technological University, Singapore 639798}
\IEEEauthorblockA{\IEEEauthorrefmark{2}Sense and Sense-abilities Programme, Institute for Infocomm Research, Singapore 138632}
\IEEEauthorblockA{\IEEEauthorrefmark{3}School of Information Systems, Singapore Management University, Singapore 188065}
}

\maketitle
\begin{abstract}
Compressive sensing has been successfully used for optimized operations in wireless sensor networks. However, raw data collected by sensors may be neither originally sparse nor easily transformed into a sparse data representation. This paper addresses the problem of transforming source data collected by sensor nodes into a sparse representation with a few nonzero elements. Our contributions that address three major issues include: 1)~an effective method that  extracts population sparsity of the data, 2)~a sparsity ratio guarantee scheme, and 3)~a customized learning algorithm of the sparsifying dictionary. We introduce an unsupervised neural network to extract an intrinsic sparse coding of the data. The sparse codes are generated at the activation of the hidden layer using a sparsity nomination constraint and a shrinking mechanism. Our analysis using real data samples shows that the proposed method outperforms conventional sparsity-inducing methods.
\end{abstract}

%\category{E.4}{Coding and Information Theory}{Data Compaction and Compression}
%\category{C.2.1}{Computer-communication Networks}{Network Architecture and Design}[wireless communication]
%\category{I.2.6}{Artificial Intelligence}{Learning}[connectionism and neural nets]
%\terms{Algorithms}

\begin{abstract}
Sparse coding, compressive sensing, sparse autoencoders, wireless sensor networks.
\end{abstract}

\section{Introduction}

A sparsely-activated data (a few nonzero elements in a sample vector) may naturally exist for compressive sensing (CS) applications in wireless sensor networks (WSNs) such as the path reconstruction problem~\cite{liu2014path}, indoor localization~\cite{fu2010compressive}, and sparse event detection~\cite{meng2009sparse}. On the other hand, a sparse data representation cannot be easily induced in many other real-world contexts (e.g., in meteorological applications and environmental data gathering). In particular, noise patterns are usually presented in collected data from WSNs which greatly affect the performance of conventional sparsity-inducing (transformation) algorithms such as the Haar wavelet and discrete cosine transforms~\cite{quer2009interplay}. This motivates the quest for noise-robust and effective sparsity-inducing methods for WSNs.

One of the breakthroughs in recent deep learning paradigms for finding high level data abstractions is achieved by introducing sparsity constraints on data representations, e.g., the Kullback\textendash{}Leibler divergence~\cite{lee2008sparse}, rectifier function\cite{glorot2011deep}, and topographic coding~\cite{lecun2012learning}. These methods are introduced for extracting intrinsic features from the data in a similar way that the human brain does while encoding sensory organ data, e.g., the low percentage of spikes in a visual cortex~\cite{lennie2003cost}. In particular, sparse deep learning methods generate sparse representations across training data for each single unit (i.e., \emph{lifetime sparsity}), and they neither guarantee sparsity for each input signal nor assert on the number of nonzero values in the sparse codes. However, a practical CS implementation in WSNs requires a sparse representation for each input signal (i.e., \emph{population sparsity}) with a sparsity ratio guarantee. Specifically, the CS solution to the underdetermined system (more number of unknowns than the number of equations) is dependent on the sparsity ratio of the signal, and the sparsity-inducing mechanism must assert an upper limit for the sparsity ratio. This sparsity bounding is necessary in WSNs as it enables using only one flat acquisition matrix for data encoding in the node. Therefore, it reduces the CS overhead in terms of memory for storing many measurement matrices in transmitting node and data control exchange as there is no need to send out rate control messages.

The main contributions of this paper can be summarized into three folds as follows.  
\begin{enumerate} 
\item This paper introduces an effective, population sparsity-inducing algorithm with sparsity ratio guarantee. The algorithm is based on a customized unsupervised neural network model of three layers (also called an autoencoder network) that generates the required, sparse coding at the second (hidden) layer. In the proposed \emph{shrinking sparse autoencoder} (SSAE), the sparsity is achieved by introducing a regularization term to the cost function of the basic autoencoder.  
\item We customize the learning algorithm to meet WSN characteristics. For example, the activations of the hidden layer during parameter learning stage are rounded to only three places to consider limited computational precision of the node. The rounding considers the low precision computations of sensor nodes, and it reduces the compressed data size and data transmission load. 
\item We present a customized learning method that optimizes the SSAE cost function. Basically, the back propagated error is only used to update the nonzero and active neurons with dominant output values for each input pattern. Moreover, a shrinking mechanism that guarantees the sparsity bound is also used during the learning of the SSAE's parameters. Therefore, an SSAE asserts on the number of nonzero elements generated at any time instant. 
\end{enumerate}

The literature is rich with sparsity-based methods that are designed for optimized WSN operations~\cite{liu2014path,fazel2013random,fu2010compressive,meng2009sparse,masiero2009data,quer2009interplay,bajwa2006compressive,luo2009compressive,griffin2007compressed,misra2012efficient}. Nonetheless, much less attention is given to the sparsity-inducing stage, and using straightforward methods to extract sparsity basis is common in previous studies such as using principal component analysis (PCA)~\cite{masiero2009data}, discrete cosine transform (DCT)~\cite{bajwa2006compressive,luo2009compressive,quer2009interplay}, discrete Fourier transform (DFT)~\cite{fazel2013random}, discrete wavelet transforms~\cite{griffin2007compressed,quer2009interplay}, and difference matrices~\cite{quer2009interplay,wu2012situ}. However, the sparse coding discipline has evolved considerable advances that significantly enhance the sparsity-inducing and hence overall WSN operations. Therefore, this paper is intended to introduce a robust and more effective sparsity-inducing method. The proposed method consists of three steps: (i)~data collection, (ii)~offline training and modeling, and (iii)~online sparse code generation. An example of the online sparse code generation for a CS application is shown in Figure~\ref{fig:example_cws} which will be described in details later.

The rest of the paper is organized as follows. In Section~\ref{sec:problem_formulation}, the problem formulation is presented. Section~\ref{sec:proposed_method} describes the proposed algorithm and the SSAE structure. Then, Section~\ref{sec:practical_considerations} discusses important practical issues of training and fitting the proposed model. In Section~\ref{sec:numerical_results}, numerical results using real-world data set are presented. Finally, Section~\ref{sec:summary} summarizes this paper.

\section{Problem Formulation}\label{sec:problem_formulation}

\begin{figure}
\begin{centering}
\includegraphics[width=0.85\columnwidth,trim=0.5cm 1cm 0.5cm 1cm]{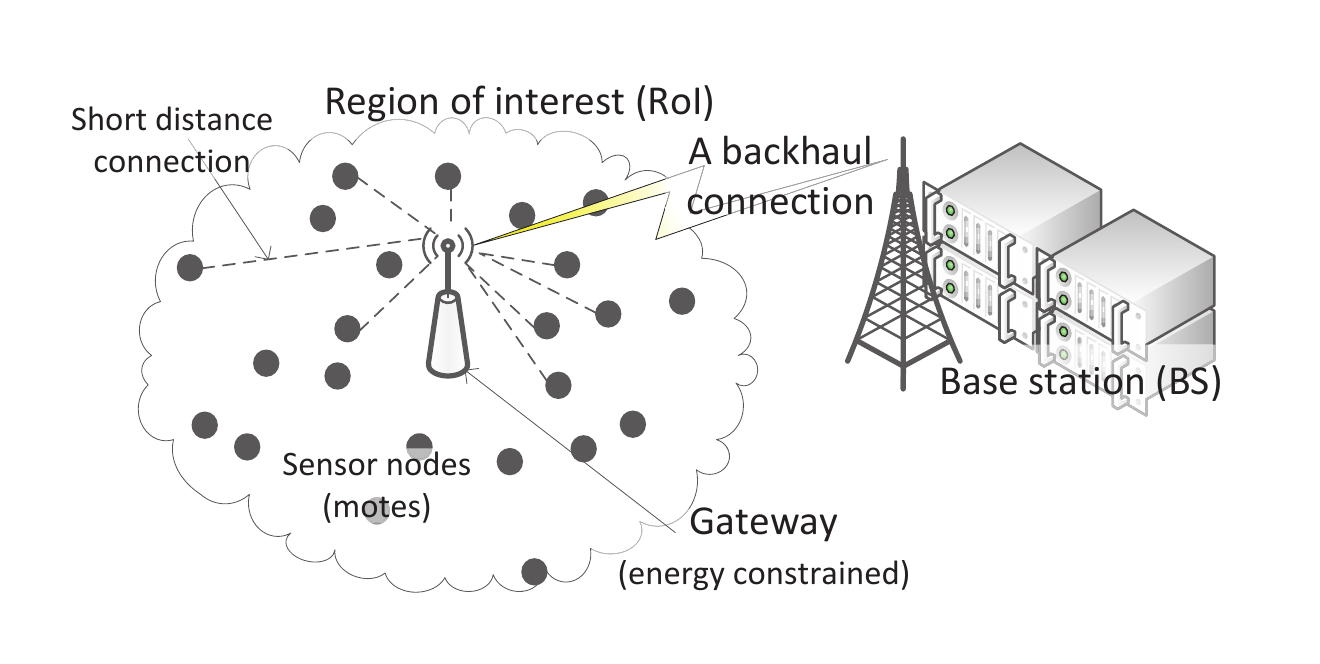}
\par\end{centering}
\caption{\label{fig:cws}Compressive sensing (CS) based data aggregation model: The RoI is assumed to be relatively far from the BS. Therefore, a gateway is designed to transmit compressed data over a costly long distance wireless connection.}
\end{figure}

Consider a dense wireless sensor network consisting of $N$ nodes, as in Figure~\ref{fig:cws}, that collects data about a region of interest (RoI). Each sensor $i$ (where $i=1,\ldots,N$) collects a real-valued sample $x_{i}$ (e.g., temperature measurements) at a predefined sampling period and transmits packets at a configured transmission power that is not sufficient to reach the base station (BS) due to long distance propagation. Therefore, a gateway (GW) is used to collect a data vector $\mathbf{x}\in\mathbb{R}^{N}$ from all sensor nodes and relay it to the BS for further analysis and processing. Thereafter, a historical data matrix $\mathbf{X\in\mathbb{R}}^{T\times N}$ is formulated at the BS containing the collected data vectors as its rows, where $T$ is the number of collected vectors. Here, the sensors' oscillators are assumed to be synchronized to the GW's clock.

After collecting sufficient historical samples (details of data collection are elucidated in Section~\ref{sub:data_collection}), and as the GW is energy and bandwidth constrained, the GW employs CS to spatially compress the data into a smaller data size. The radio transceiver is the most energy consuming unit in an ordinary sensor node~\cite{pottie2000wireless}. Thereby, the energy consumption becomes more critical in the GW unit as it transmits huge data over the backhaul connection, while sensor nodes are assumed to transmit for short distances. It is important to note that our algorithm can be also temporally applied at each individual sensor node. However, data delivery latency is provoked as temporal samples must be collected at the node before being transmitted as one compressed chunk. Next, we give an overview of the CS framework and its implementation at the GW device, and the data reconstruction at the BS unit.

\subsection{Compressive Sensing (CS)}

CS is a signal processing method for effective data recovery from a few data samples than the Nyquist rate~\cite{qaisar2013compressive}. Assuming a sparse signal $\mathbf{s}\in\mathbb{R}^{L}$ that has only $K$ nonzero elements; therefore, $\mathbf{s}$ is called a $K$-sparse signal, and the \emph{sparsity ratio} $\eta$ is equal to $\frac{K}{L}$. Moreover, suppose a measurement (or sensing) matrix $\boldsymbol{\Phi}\in\mathbb{R}^{M\times L}$ that obeys the restricted isometry property (RIP)~\cite{candes2008restricted}. Here, $M$ is assumed to be much smaller than $L$; therefore, $\boldsymbol{\Phi}$ is a flat matrix with more columns than rows. The sensing system under consideration that is executed by the GW to compress data can be expressed as
\begin{equation}
\mathbf{y}=\boldsymbol{\Phi}\mathbf{s}
\end{equation}
where $\mathbf{y}\in\mathbf{\mathbb{R}^{M}}$ is the resulted measurement vector. $\boldsymbol{\Phi}$ can be sampled from different distributions to meet the RIP such as the Gaussian distribution~\cite{chen2005condition}. Moreover, for high probability recovery, $M$ must also meet the following constraint~\cite{candes2006robust}:
\begin{equation}
M\geq\rho K\log_{2}\left(\frac{L}{K}\right)\label{eq:cs_measurements}
\end{equation}
where $\rho$ is a constant, and $M\ll N$. At the BS unit, the reconstruction of $\mathbf{s}$ from $\mathbf{y}$ can be achieved by minimizing the following relaxed problem~\cite{candes2006stable}:
\begin{equation}
\mathbf{s}^{*}=\arg\min_{\left\Vert \mathbf{y}-\boldsymbol{\Phi}\mathbf{s}\right\Vert _{2}\leq\epsilon}\left\Vert \mathbf{s}\right\Vert _{1}\label{eq:cs_reconstruction}
\end{equation}
where $\epsilon$ is a small constant. The optimization problem (\ref{eq:cs_reconstruction}) can be solved using a regularized least square method called least absolute shrinking and selection operator (LASSO)~\cite{tibshirani1996regression}.

\subsection{Sparsity-inducing}

Clearly, the whole CS framework is based on the sparsity assumption. Natural signal such as sound and images can be transformed into a sparse form by projecting them into a suitable basis~\cite{qaisar2013compressive}. However, this is not the case when dealing with WSN data. More precisely, sensor nodes produce noisy readings of the form
\begin{equation}
\mathbf{x}=\mathbf{x}^{*}+\mathbf{z}
\end{equation}
where $\mathbf{x}^{*}\in\mathbb{R}^{N}$ is the noiseless data vector of the physical phenomenon, and $\mathbf{z}\in\mathbb{R}^{N}$ is the added noise vector. Noise values are assumed to be independent Gaussian variables with zero mean and variance $\sigma_{z}^{2}$ such that $z\sim\mathbb{N}\left(0,\sigma_{z}^{2}I_{N}\right)$. Therefore, even through the neighbor sensors are spatially correlated and hence compressible, the noise existence hampers the accurate approximation of source signal $\mathbf{x}$ using linear projection methods. In particular, smooth signal are representable using linear combinations of Fourier bases, and smooth piecewise signals are linearly representable in wavelet bases~\cite{bajwa2006compressive}. Nonetheless, the smoothness condition is not guaranteed in sensor data as data samples are usually affected by noise patterns, and commercial sensors sense phenomenon with finite precision and are not noise robust. For example, a few noise readings can destroy the sparsity pattern of a DCT transformed data~\cite{luo2009compressive}. 

The main aim of any robust sparsity-inducing mechanism is to transform the source signal $\mathbf{x}\in\mathbb{R}^{N}$ into the sparse signal $\mathbf{s\in\mathbb{R}}^{L}$. An upper bound guarantee on the sparsity ratio of the generated signal $\mathbf{s}$ is a ``must-have'' feature in most sparsity-based applications such as in CS. In particular, this guarantee enables designing low memory and low communication overhead applications for WSNs as a single sensing matrix $\boldsymbol{\Phi}$ is used by the GW unit to compress data. Then, the BS does not require any information from the GW to recover the reconstruction signal $\overline{\mathbf{x}}$ other than the measurement vector $\mathbf{y}$, where $\overline{\mathbf{x}}$ is a reconstruction of the noiseless data vector $\mathbf{x}^{*}$.

\begin{figure}
\begin{centering}
\includegraphics[width=0.85\columnwidth,trim=0.5cm 1.2cm 0.5cm 0.5cm]{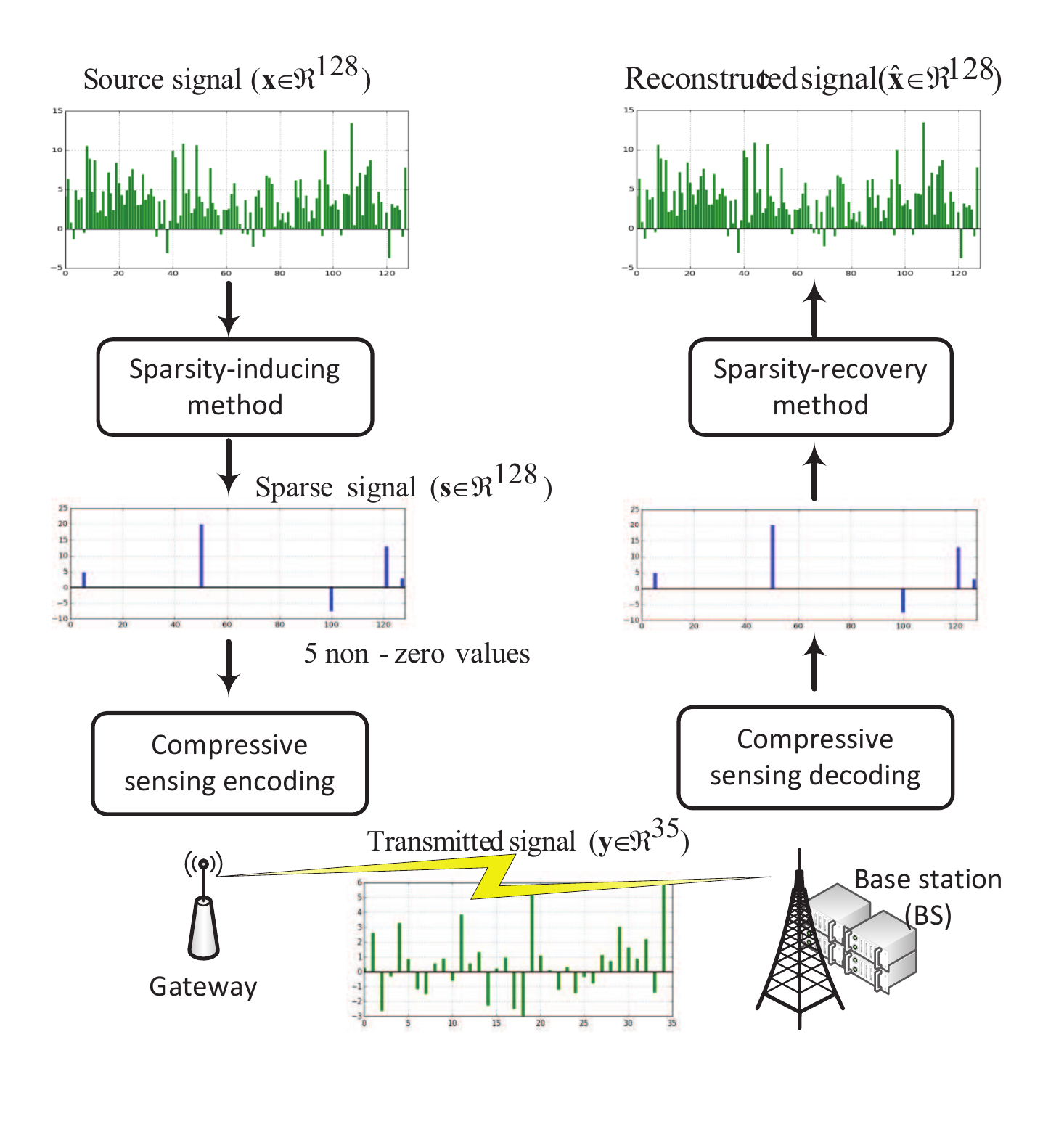}
\par\end{centering}
\caption{\label{fig:example_cws}Example of data compression, transmission, and recovery operations using CS and sparsity-inducing models. }
\end{figure}

An example of the system online operational procedure is shown in Figure~\ref{fig:example_cws} which includes the sparsity-inducing and CS components. The next section presents the proposed sparsity-inducing mechanism.

%%%%%%%%%%%%%%%%%%%%%%%%%%%%%%%%%%%%%%%%%%%%%%%%%%%%%%%%%%%%%%%%%%%%%%%%%%%%%%%%%%%%%%%%%%%%%%%
\section{Shrinking Sparse Autoencoder (SSAE)}\label{sec:proposed_method}

\begin{figure}
\begin{centering}
\includegraphics[width=0.75\columnwidth,trim=0.5cm 1.2cm 0.5cm 0.5cm]{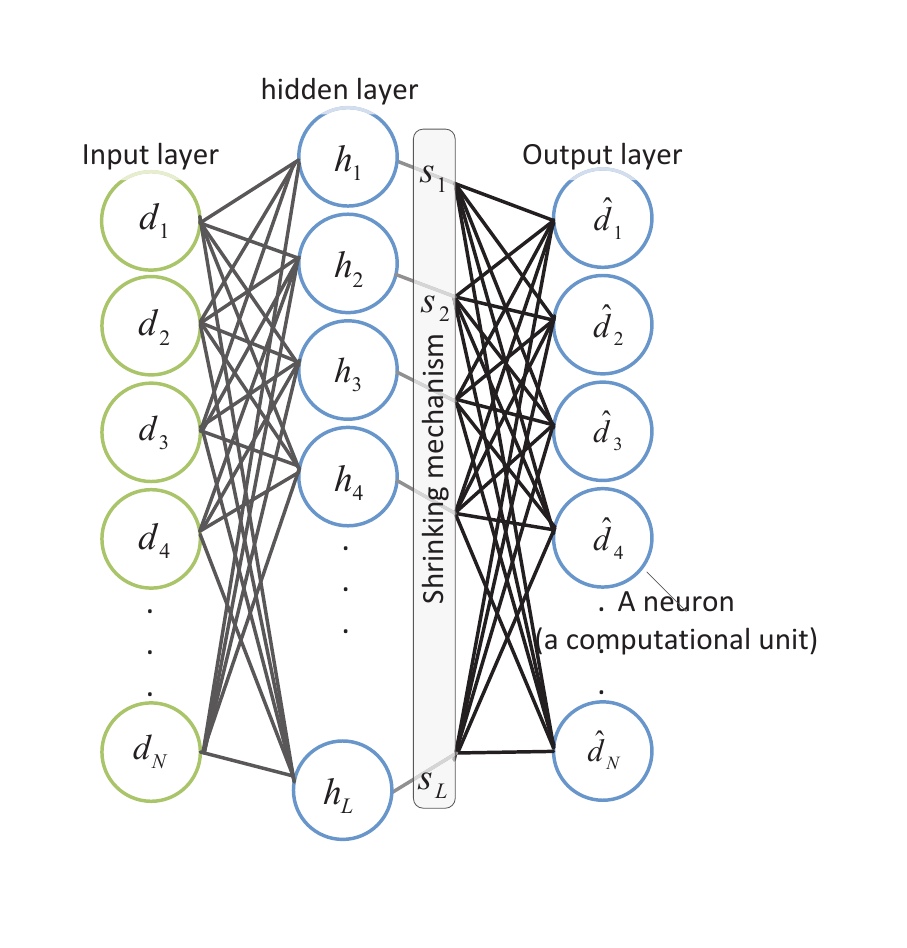}
\par\end{centering}
\caption{\label{fig:ssae}Illustration of the SSAE structure.}
\end{figure}

In this section, we introduce an autoencoder's variant which we call shrinking sparse autoencoder (SSAE) as shown in Figure~\ref{fig:ssae}. The SSAE network is specially designed to transform sensory data from its original domain into a sparse domain. The SSAE structure consists of three neural (or computational unit) layers. Firstly, an input layer that is connected to the input signal $\mathbf{d\in\mathbb{R}}^{N}$, where $N$ is the number of sensor nodes in the network. Briefly, $\mathbf{d}$ is a sphered version of the raw sensor data $\mathbf{x}$, where $\left\{ d_{i}\in\mathbb{R}|-1<d_{i}<1\right\} $ as described in Section~\ref{sub:sphering}. Secondly, a hidden layer is used to generate an intrinsic code $\mathbf{h}\in\mathbb{R}^{L}$ at its activation. Thirdly, an output layer that includes the same number of neurons as the input layer and generates a recovery of the input data $\hat{\mathbf{d}}\in\mathbb{R}^{N}$. The layers are connected to each other using the following formulations:
\begin{eqnarray}
\mathbf{h} & = & f\left(\mathbf{W}^{(1)}\mathbf{d}+\mathbf{b}^{(1)}\right)\label{eq:hidden_activation}\\
\hat{\mathbf{d}} & = & f\left(\mathbf{W}^{(2)}\mathbf{s}+\mathbf{b}^{(2)}\right)\label{eq:output_activation}
\end{eqnarray}
where $\mathbf{W}^{(1)}$ is the weight matrix connecting the input and hidden layers, $\mathbf{W}^{(2)}$ is the weight matrix connecting the hidden and output layers, and $\mathbf{b}^{(1)}$ and $\mathbf{b}^{(2)}$ are the biases of the input and hidden layers, respectively. Additionally, $\mathbf{s}$ is the sparse data representation that is obtained by applying the shrinking operation as described in Section~\ref{sub:shrinking_operation}. For simplicity, we define $\boldsymbol{\theta}$ to contain all the SSAE's parameters such that $\boldsymbol{\theta}\doteq\left[\mathbf{W}^{(1)},\mathbf{W}^{(2)},\mathbf{b}^{(1)},\mathbf{b}^{(2)}\right]$. Moreover, $f\left(\cdot\right)$ is the non-linear hyperbolic tangent function.

The SSAE's cost function $\Gamma\left(\cdot\right)$ includes two terms as follows:
\begin{multline}
\Gamma\left(\boldsymbol{\theta};\mathbf{D}\right) = \frac{1}{T}\left(\sum_{u=1}^{T}\frac{1}{2}\left\Vert \widehat{\mathbf{d}}^{(u)}-\mathbf{d}^{(u)}\right\Vert ^{2}\right) + \\
 \frac{\gamma}{T}\left(\sum_{u=1}^{T}\log_{10}\left(1+\left(\mathbf{h}^{(u)}\right)^{2}\right)\right)
\label{eq:cost_function}
\end{multline}
where $\mathbf{D\in\mathbb{R}}^{T\times N}$ is the training matrix of historical data such that each input vector $\left\{ \mathbf{d}^{(u)}\right\} _{u=1}^{T}$ is stored in a row of this matrix, and $\mathbf{h}^{(u)}$ is the hidden layer activation of $\mathbf{d}^{(u)}$. Moreover, $T$ is the training set size configured at the offline training algorithm (the details are given in Section~\ref{sub:training}). As with any other autoencoder, the first term is the average sum of the difference between input vectors and their reconstructions at the output layer. This term is used to encourage the neural network to reconstruct its input data at the output layer. The second term is used to encourage sparsity at the generated coding in the hidden layer. The \emph{sparsity penalty} $\gamma$ is a hyper-parameter to manage the weights of each term in the optimization problem. In other words, using a big value for $\gamma$ results in highly sparse representation, but with poor reconstruction capability. Then, the well-known delta rule can be used to update the SSAE's weights and biases as follows~\cite{rumelhart1988learning}:
\begin{eqnarray}
W^{(q)}_{ij} & \doteq & W^{(q)}_{ij}-\alpha\frac{\partial}{\partial W^{(q)}_{ij}}\Gamma\left(\boldsymbol{\theta};\mathbf{D}\right)  ,	\label{eq:w_update1}\\
b^{(q)}_{i} & \doteq & b^{(q)}_{i}-\alpha\frac{\partial}{\partial b^{(q)}_{i}}\Gamma\left(\boldsymbol{\theta};\mathbf{D}\right)\label{eq:b_update}
\end{eqnarray}
where $\alpha$ is the \emph{learning rate}, and $q\in\left\{1,2\right\}$ is the layer number within the SSAE network. These update rules are executed at each iteration of a gradient descent method. The partial derivative is given by 
\begin{equation}
\frac{\partial}{\partial W^{(q)}_{ij}}\Gamma\left(\boldsymbol{\theta};\mathbf{D}\right)=\frac{1}{T}\sum_{u=1}^{T}\frac{\partial}{\partial W^{(q)}_{ij}}\Gamma\left(\boldsymbol{\theta};\mathbf{d}^{(u)}\right)\label{eq:partial_derivative}
\end{equation}
where $\Gamma\left(\boldsymbol{\theta};\mathbf{d}^{(u)}\right)$ is the cost function defined for a single sample $\mathbf{d}^{(u)} \in \mathbf{D}$. This means that the overall partial derivative of (\ref{eq:cost_function}) is found by averaging the partial derivatives of all input samples.  The second term of (\ref{eq:cost_function}) only affects the partial derivative of the hidden layer ($q=2$) which is computed as follows:
\begin{multline}
\frac{\gamma}{T}\sum_{u=1}^{T}\frac{\partial}{\partial W^{(2)}_{ij}}\left(\gamma\log_{10}\left(1+\left(\mathbf{h}^{(u)}\right)^{2}\right)\right)= \\
\frac{\gamma}{\log_{e}\left(10\right)\times T}\sum_{u=1}^{T}\left(\frac{2\mathbf{h}^{(u)}}{1+\left(\mathbf{h}^{(u)}\right)^{2}}\right)f'\left(\mathbf{W}^{(2)}\mathbf{d}^{(u)}+\mathbf{b}^{(2)}\right)
\end{multline}
where $f'\left(\cdot\right)=1-\left(f\left(\cdot\right)\right)^2$ is the element-wise derivative of $f\left(\cdot\right)$. Thereby, the SSAE is designed to generate many zeros at the hidden layer. One can think of a neuron as being active when its output is not equal to zero, and an inactive neuron does not participate in forwarding the input data to the output (because it does not generates signals). To this end, two important issues of the second term of (\ref{eq:cost_function}) must be noted as follows:
\begin{itemize} 
\item The second term minimizes the hidden layer activation, but it still does not ensure exactly zero activations. 
\item It does not guarantee sparsity ratio at the generated codes. 
\end{itemize} 
Accordingly, a shrinking mechanism must be applied at the hidden layer activation and before propagating them to the output layer to reconstruct the input. In particular, one can think of the second term as only being used as a mechanism of nominating the most promising neurons to be deactivated by the shrinking mechanism as described in the next section.

\subsection{Shrinking (Pruning) Scheme}\label{sub:shrinking_operation}

\begin{algorithm}
\begin{algorithmic}[1]

\State \textbf{Input} $\mathbf{h}\in\mathbb{R}^{L}$: hidden layer activation before shrinking

\State \textbf{Input} $K$: maximum nonzero activations

\State $\mathbf{s}=\mathbf{h}$\Comment{copy operation }

\For{$i=0$ to $L-K$}

\State $p=0$

\For{$j=0$ to $(L-1$)}

\If{$\left|s_{p}\right|>\left|s_{j}\right|$ \textbf{and} $\left|s_{j}\right|>0$}

\State $p=j$

\EndIf

\EndFor

\State $s_{p}=0$\Comment{zero-out smallest value}

\EndFor

\State \textbf{Output} $\mathbf{s}\in\mathbb{R}^{L}$

\end{algorithmic} 

\caption{\label{alg:shrinkage}Pseudo-code for the shrinking operation of hidden layer's neurons.}
\end{algorithm}

Even though the cost function of the SSAE is designed to generate a sparse data coding at the hidden layer, it does still neither guarantee a coding with population sparsity (sparsity at each input vector) nor assert on the maximum nonzeros for each input. Equally important, it will most likely generate values close to, but not absolutely zero. Therefore, we propose a simple shrinking mechanism that can complete the design cycle. For each input vector, the proposed shrinking mechanism ``zero out'' the least dominant neurons from the hidden layer, and therefore switching them to the deactivation mode. The least dominant neurons are the ones with the least effect on the data reconstruction at the output layer, and hence the minimum activation values that result from the sparsity restrictions. Therefore, only $K$ active neurons at the hidden layer  forward propagate the input through the SSAE network, and the remaining $L-K$ neurons are switched off. An optimized implementation of the shrinking scheme is given by the pseudo-code in Algorithm~\ref{alg:shrinkage}, where $\left|\cdot\right|$ is the absolute value function.

\subsection{Offline Training}\label{sub:training}

The SSAE's parameter adjustment is done during an offline training stage. As a resource demanding process, the training must be performed at the BS unit, and then the SSAE's parameters ($\boldsymbol{\theta}$) are disseminated for online data compression at GW. The learning stage and SSAE's parameters are tuned using a resourceful BS with relatively high precision operations. However, GW is usually constrained in terms of computational resources and computational precision (i.e., the machine epsilon value). Therefore, rounding the activation at the hidden layer is needed during the learning stage to match the GW's low precision. Moreover, with rounding, less data is transmitted from GW to BS.

To learn the SSAE's parameters ($\boldsymbol{\theta}$), we minimize (\ref{eq:cost_function}) by using a non-linear quasi-Newton method called the limited-memory Broyden-Fletcher-Goldfarb-Shanno (L-BFGS) method~\cite{byrd1995limited}. However, firstly the collected historical data $\mathbf{X}\in\mathbb{R}^{T\times N}$ must be randomly shuffled. This is because sensors' readings are highly correlated over time, and a non-shuffled data causes the SSAE to dominantly learn the training data' patterns in training data only. Therefore, the shuffling step ensures that the training and testing data sets contain all possible data patterns.  Moreover, the cross validation technique~\cite{kohavi1995study} is an effective method for testing the model generalization capabilities, while benefiting from all available samples for training. Cross validation divides the training data into $\varphi$ groups (e.g., 10 groups) then at each time, one group is held out for testing while using the remaining for model fitting. Then, the model performance is found by averaging errors of all cross validation's groups. The offline learning is described in Algorithm~\ref{alg:training}.

The learning algorithm is computationally intensive for sensor nodes and must be performed at the BS. Moreover, if the statistical parameters of the underlying phenomenon change, the offline training must be re-executed and an updated $\left[\mathbf{W}^{(1)},\mathbf{b}^{(1)}\right]$ should be disseminated to the nodes.

\begin{algorithm}
\begin{algorithmic}[1]

\State \textbf{Input} $\mathbf{X}\in\mathbb{R}^{T\times N}$: historical sensor data

\State \textbf{Input} $K$: maximum nonzero activations

\State \textbf{Input} $\gamma$: sparsity hyper-parameter

\State \textbf{Input} $\varphi$: number of folds for cross validation

\State Randomly shuffle $\mathbf{X}$

\State Divide $\mathbf{X}$ into $\varphi$ folds $\mathbf{X}^{1},\ldots,\mathbf{X}^{\varphi}$

\ForAll{$\mathbf{X}^{i},i=1,\ldots,\varphi$}

\ForAll{$\mathbf{x}\in(\mathbf{X}\setminus\mathbf{X}^{i})$}\Comment{held out $\mathbf{X}^{i}$ for testing}

\State Sphere input $\mathbf{x}$ to get $\mathbf{d}$ using (\ref{eq:sphere})

\State Append $\mathbf{d}$ to $\mathbf{D}$

\EndFor

\Repeat

\ForAll{$\mathbf{d}\in\mathbf{D}$}

\State Forwardly propagate $\mathbf{d}$ to compute $\mathbf{h}$
using (\ref{eq:hidden_activation})

\State Shrink $\mathbf{h}$ to get $\mathbf{s}$ as in Algorithm~\ref{alg:shrinkage}

\State Compute $\hat{\mathbf{d}}$ using (\ref{eq:output_activation})

\EndFor

\State Compute the cost value using (\ref{eq:cost_function})

\State Compute the gradient vector as in (\ref{eq:partial_derivative})

\State Update $\boldsymbol{\theta}$ using the L-BFGS method

\Until{ learning converges}

\State Compute accuracy using $\mathbf{X}^{i}$

\EndFor

\State Compute average accuracy of the $\varphi$ folds

\State \textbf{Output} $\boldsymbol{\theta}\doteq\left[\mathbf{W}^{(1)},\mathbf{W}^{(2)},\mathbf{b}^{(1)},\mathbf{b}^{(2)}\right]$

\end{algorithmic} 

\caption{\label{alg:training}The offline training algorithm.}
\end{algorithm}

\subsection{Computational Complexity}
The online encoding and decoding of sparse codes are lightweight. In particular, the GW (or a sensor node) can generate sparse codes by only using $\left[\mathbf{W}^{(1)},\mathbf{b}^{(1)}\right]$ as in (\ref{eq:hidden_activation}) and Algorithm~\ref{alg:shrinkage} with $\ensuremath{\mathcal{O}(L\times N)}$ of overall time complexity. The data recovery is performed at the BS by using $\left[\mathbf{W}^{(2)},\mathbf{b}^{(2)}\right]$ as in (\ref{eq:output_activation}) with a similar time complexity of $\ensuremath{\mathcal{O}(L\times N)}$. 

\subsection{Sparse Codes}

\begin{figure}
\begin{centering}
\subfloat[\label{fig:learning_curve}]{\begin{centering}
\includegraphics[width=0.7\columnwidth,trim=1cm 0.7cm 1cm 0cm]{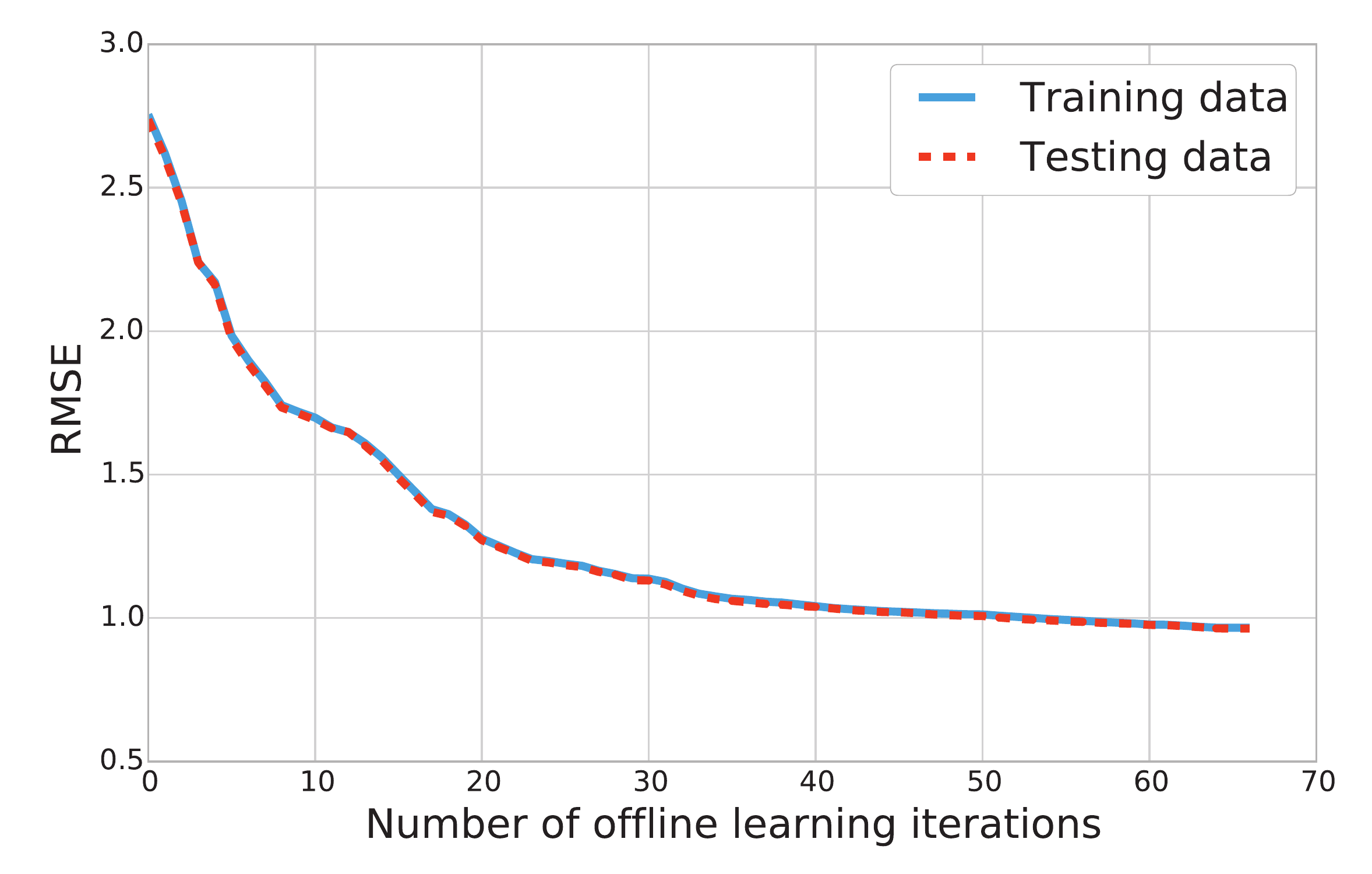}
\par\end{centering}
}
\par\end{centering}
\begin{centering}
\subfloat[\label{fig:population_sparsity}]{\begin{centering}
\includegraphics[width=0.85\columnwidth,trim=0.5cm 0.7cm 2cm 1cm]{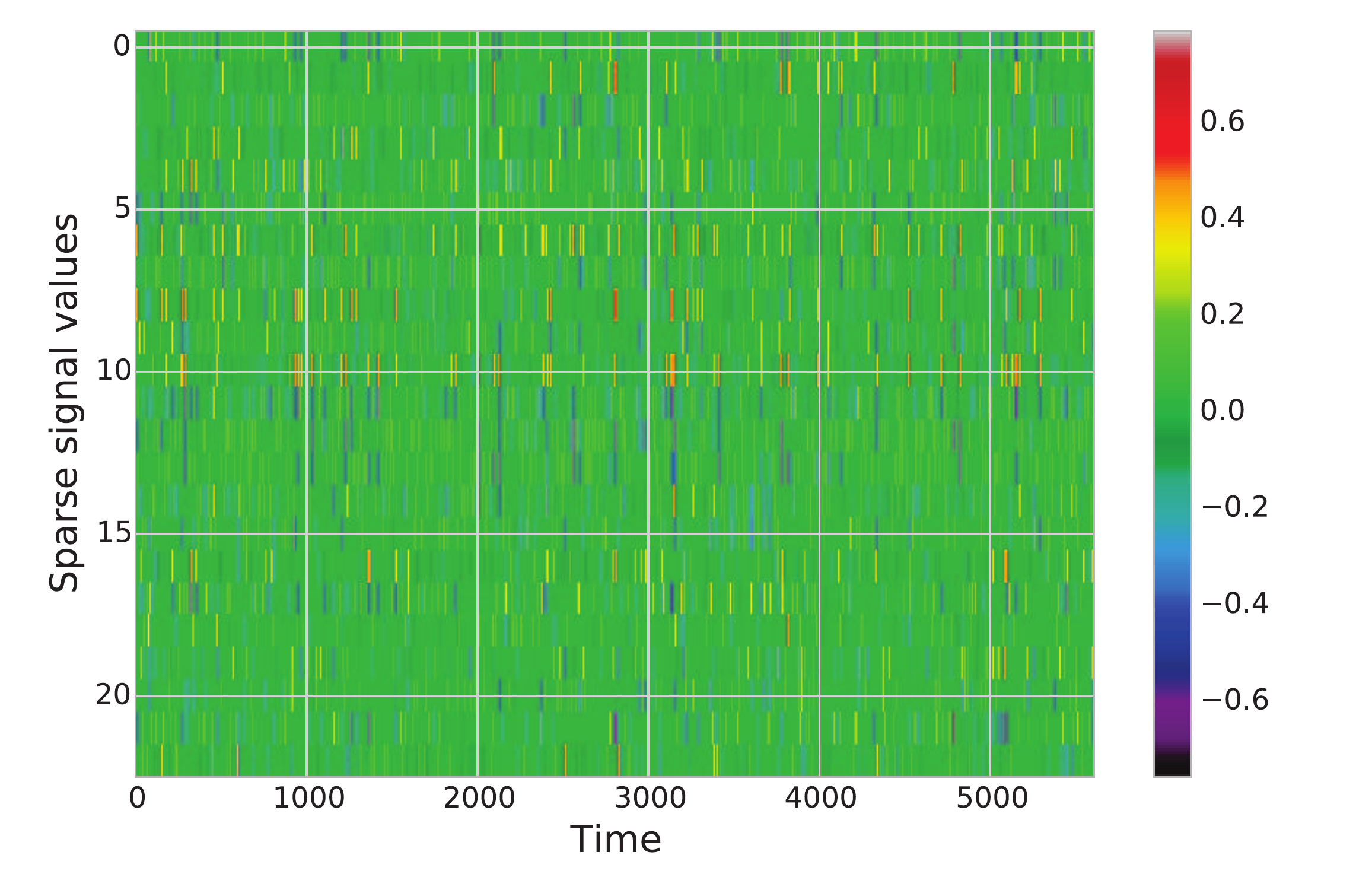}
\par\end{centering}
}
\par\end{centering}

\caption{An SSAE which is designed to produce a maximum of 5 nonzero values at each time instant ($\eta=0.2$) for 23 sensors (i.e., $\mathbf{x}\in\mathbb{R}^{23}$). (a)~A learning curve that shows the convergence of the offline learning algorithm, and (b)~activation values of the hidden layer's neurons.}
\end{figure}

For the verification and analysis in the following sections, a meteorological data set from the Sensorscope project~\cite{Sensorscope_Grand} is used. The data set contains surface temperature samples of 23 sensors. The learning curve of the SSAE is shown in Figure~\ref{fig:learning_curve}.

An important indication of successful SSAE training is ensuring that hidden neurons are not connected with zero weights to the input layer. In other words, this ensures that any neuron in the hidden layer will be active for some input patterns, and hence no ``always-off'' neuron exists. This increases the model performance of generalizing to non local data, and hence it performs well on extremely non linear data, as all neurons participating increases the possible code formulations (i.e., the number of distinct combinations is increased when having more active neurons). Figure~\ref{fig:population_sparsity} shows hidden layer activations over time. Here, two main desirable properties can be observed 
\begin{enumerate} 
\item Population sparsity is achieved, and the maximum number of active neurons at any time instant is guaranteed by the SSAE network. This upper bound of nonzeros in a generated sparse code considers the tradeoff between the recovery error and compression ratio of the data aggregation model. Therefore, only a single sensing matrix is needed when using CS to create a measurement vector at the GW node. 
\item All neurons are participating in the sparse code generation, and without any ``always-off'' neuron. Moreover, the activation values of the active neurons are not concentrated around very small values near zero. This feature cannot be achieved in conventional average activity ratio sparse autoencoders, such as the Kullback\textendash{}Leibler divergence, as they are designed for lifetime sparsity only.
\end{enumerate}

%\begin{figure}
%\begin{centering}
%\includegraphics[width=0.75\columnwidth,trim=1cm 0.8cm 5cm 1cm]{Figures/population_sparseness}
%\par\end{centering}
%\caption{\label{fig:population_sparsity}Activation values of the hidden layer's neurons, where the SSAE is required to produce a maximum of 5 nonzero values at each %time instant ($\eta=0.2$). }
%\end{figure}

%%%%%%%%%%%%%%%%%%%%%%%%%%%%%%%%%%%%%%%%%%%%%%%%%%%%%%%%%%%%%%%%%%%%%%%%%%%%%%%%%%%%%%%%%%%%%%%
\section{Discussion and Practical Considerations}\label{sec:practical_considerations}

In this section, some practical issues of the SSAE training and fitting are discussed.

\subsection{Data Collection}\label{sub:data_collection}

\begin{figure}
\begin{centering}
\includegraphics[width=0.6\columnwidth,trim=1.2cm 0.6cm 1.2cm 0.5cm]{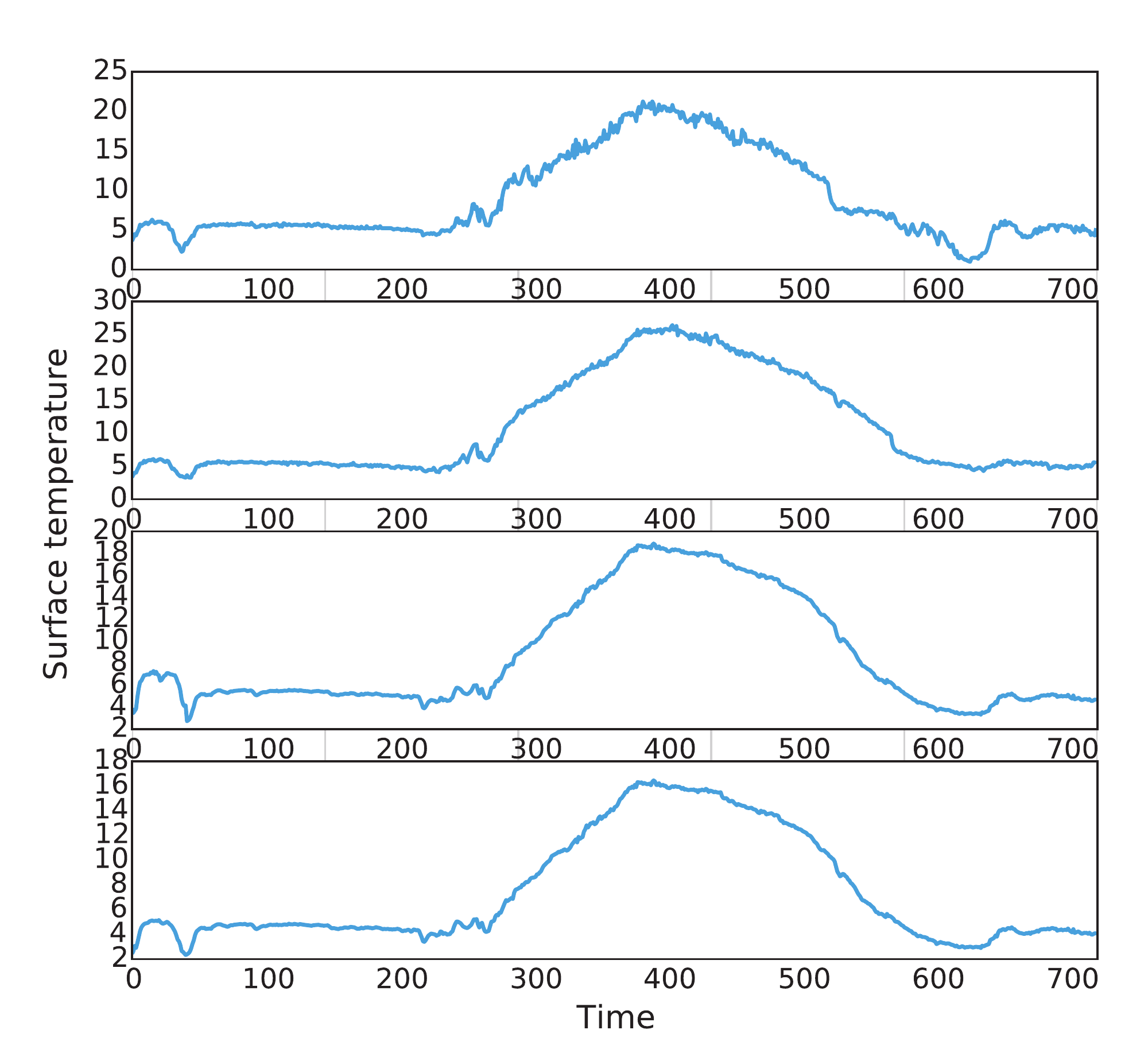}
\par\end{centering}
\caption{\label{fig:spatial_correlation}Surface temperature readings of 4 neighbor sensors from the Sensorscope deployment over 1 day (1 sample every 2 minutes). This shows the spatial correlation among sensors' measurements, and hence data compressibility.}
\end{figure}

A crucial aspect of machine leaning-based approaches, such as the SSAE network, is the training data requirement. A system designer may have access to a large historical data set that is collected in the past. This historical data can be used to train the SSAE's model. However, this is not the case for new WSN's deployments, and the lack of sensor data hinders the accurate fitting of the SSAE's parameters (i.e., $\boldsymbol{\theta}$). Clearly, the SSAE's model needs to globally generalize to unseen data samples. In any machine learning method, having more training data can improve generalization performance, but having more data is not the only solution~\cite{domingos2012few}. In WSNs, the following issues must be considered when using an SSAE as a sparsity inducing method. 
\begin{enumerate} 
\item It is assumed that sensor nodes are densely deployed and hence spatially correlated with each other (e.g., as in Figure~\ref{fig:spatial_correlation} for the Sensorscope project's data). SSAE learns these spatial correlation and redundant patterns in the nodes' collected data. Therefore, if the underlying phenomenon becomes different in the way that it changes the nodes' spatial correlation, then new data collection and offline model fitting must be performed. 
\item The amount of data required to fit the SSAE's model depends on the underlying sensed phenomenon, and for more complex correlation patterns among sensors, more data samples are needed.
\end{enumerate}

\subsection{Data Sphering}\label{sub:sphering}

\begin{figure}
\begin{centering}
\subfloat[]{\begin{centering}
\includegraphics[width=0.465\columnwidth,trim=1.2cm 0.7cm 1cm 1cm]{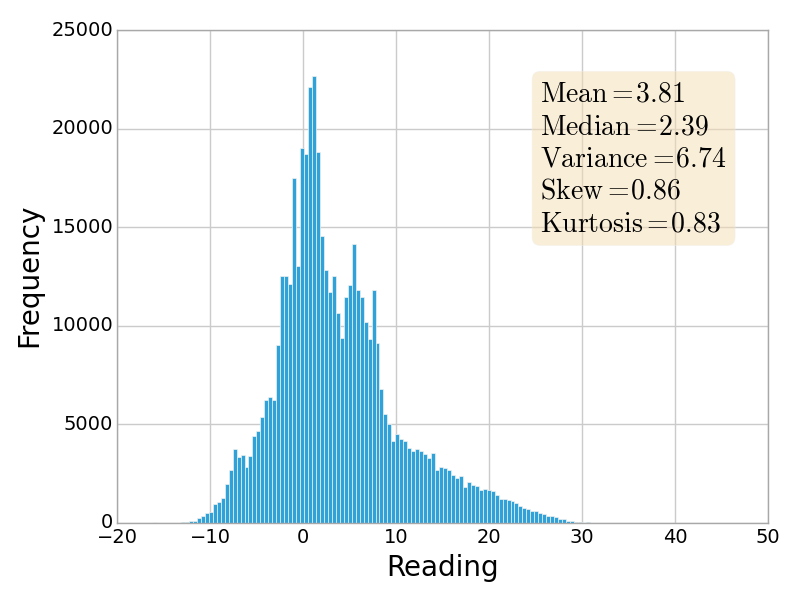}
\par\end{centering}

}\subfloat[]{\begin{centering}
\includegraphics[width=0.465\columnwidth,trim=1.2cm 0.7cm 1cm 1cm]{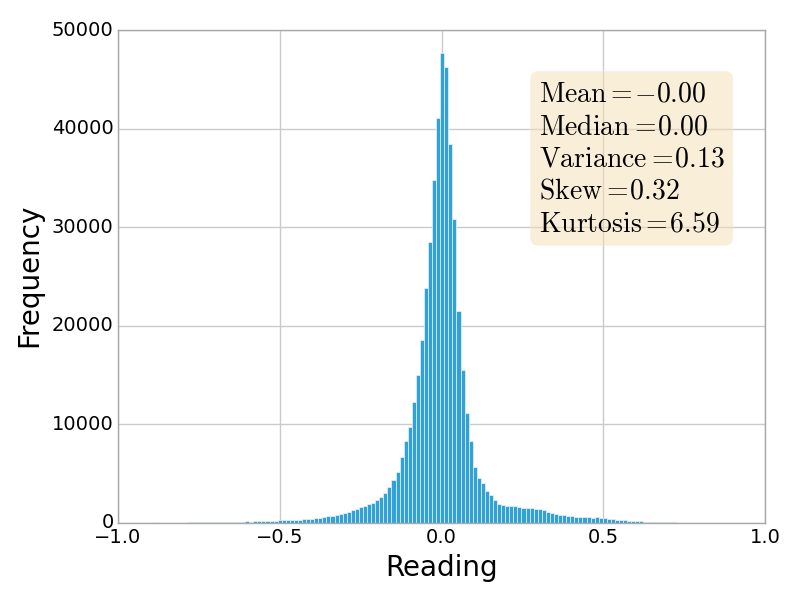}
\par\end{centering}

}
\par\end{centering}

\caption{\label{fig:data_sphering}Data sphering and its effects on data by showing histograms and basic statistical values. (a)~Raw data in the range of $[-16.15,47.91]$. (b)~Sphered data that is scaled to a new range of $(-1,1)$ with a Gaussian-like distribution.}
\end{figure}

Before using historical sensor data to train the SSAE, a pre-processing operation is required, namely the data sphering stage. Data sphering is simply achieved by applying the following operation on each sensors' raw input vector $\mathbf{x}\in\mathbb{R}^{N}$
\begin{equation}
\mathbf{d}=\textrm{sphere}(\mathbf{x},\sigma)=\frac{\max\left(\min\left(\left(\mathbf{x}-\overline{x}\right),3\sigma\right),-3\sigma\right)}{3\sigma}\label{eq:sphere}
\end{equation}
where $\sigma$ is the standard deviation of the historical training matrix $\mathbf{X}$, $\overline{x}=\frac{1}{N}\sum_{i=1}^{N}x_{i}$ is the arithmetic mean of each input vector, and again $\mathbf{d}\in\mathbb{R}^{N}$ is the SSAE's input vector which is the resulting data vector after sphering. Unlike the standard element-wise standardization, this subtracts the arithmetic mean of each input vector and not the whole training matrix's mean value. The effect of data sphering on training data is shown in Figure~\ref{fig:data_sphering}. Clearly, the data is transformed into a smoother Gaussian-like curve with zero mean (other statistical parameters are also shown). Equally important, the resulting scale of sphered data is in the $(-1,1)$ interval, which makes it suitable for the operation of the hyperbolic tangent function. In particular, the hyperbolic tangent function generates an output in the range of $(-1,1)$ and without data pre-processing to this range, the SSAE cannot produce outputs similar to input data.

The reverse operation of data sphering is required at BS to reconstruct the original raw sensors' vector $\mathbf{\hat{x}}\in\mathbb{R}^{N}$ from the SSAE's output values $\hat{\mathbf{d}}\in\mathbb{R}^{N}$. The reverse operation is given as
\begin{equation}
\mathbf{\hat{x}}=\textrm{desphere}(\hat{\mathbf{d}},\overline{x},\sigma)=3\sigma\hat{\mathbf{d}}+\overline{x}.
\end{equation}
Here, $\sigma$ is constant for all recovered vectors, and therefore can be stored at the BS. However, $\overline{x}$ must be sent from the GW to the BS along with the compressed data. Therefore, the transmitted data size when using CS is $M+1
$.

%%%%%%%%%%%%%%%%%%%%%%%%%%%%%%%%%%%%%%%%%%%%%%%%%%%%%%%%%%%%%%%%%%%%%%%%%%%%%%%%%%%%%%%%%%%%%%%
\section{Numerical Results}\label{sec:numerical_results}

In this section, we evaluate the performance of the SSAE-based sparsity inducing method.

\subsection{SSAE Tunning}\label{sub:parameter_settings}

\begin{figure}
\begin{centering}
\includegraphics[width=0.65\columnwidth,trim=1cm 0.7cm 1cm 1cm]{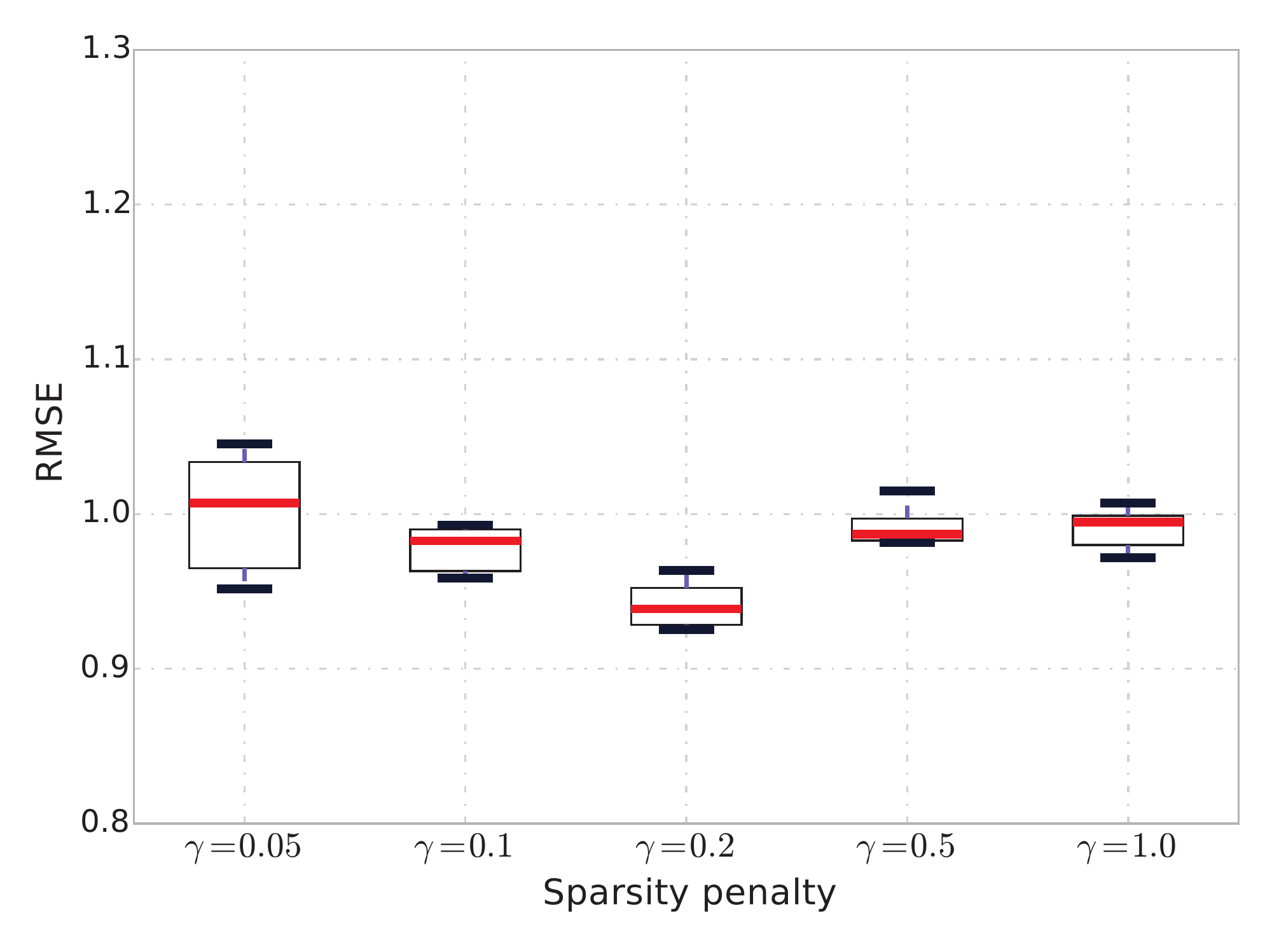}
\par\end{centering}
\caption{\label{fig:sparsity_hyperparameter}Sparsity parameter setting for $\eta=0.217$. Bars represent root mean square error (RMSE) values over 10 runs. The maximum performance achieved at $\gamma=0.2$.}
\end{figure}

One of the main difficulties of applying neural network-based methods is the numerical tuning of the network hyper-parameters. Hyper-parameter setting of autoencoder's variants can be facilitated by searching over a scale of values in the log-domain (e.g., values such as $10^{-1},10^{-2},10^{-3},\ldots$), and then the value that minimizes the cross validation error is selected accordingly~\cite{bengio2012practical}. Figure~\ref{fig:sparsity_hyperparameter} shows the setting of the sparsity hyper-parameter $\gamma$ for sparsity ratio $\eta=0.39$. The sparsity term in (\ref{eq:cost_function}) can be interpreted as sparsity nomination term, that is fed to the shrinking mechanism to generate sparse codes. Therefore, trying different values of $\gamma$ is useful to achieve maximum signal reconstruction performance. For SSAE in the next experiments, the following function is used for the sparsity penalty $\gamma$ settings
\begin{equation}
\gamma\left(\eta\right)=0.26-0.26\eta,
\end{equation}
which is found by manually fitting the hyper-parameter $\gamma$ for two values of $\eta$, as described above, and then finding the line connecting these two manually fitted points.

\subsection{Comparing to Benchmarks}\label{sub:recovery_error}

\begin{figure}
\begin{centering}
\includegraphics[width=0.75\columnwidth,trim=1.4cm 0.7cm 1cm 1cm]{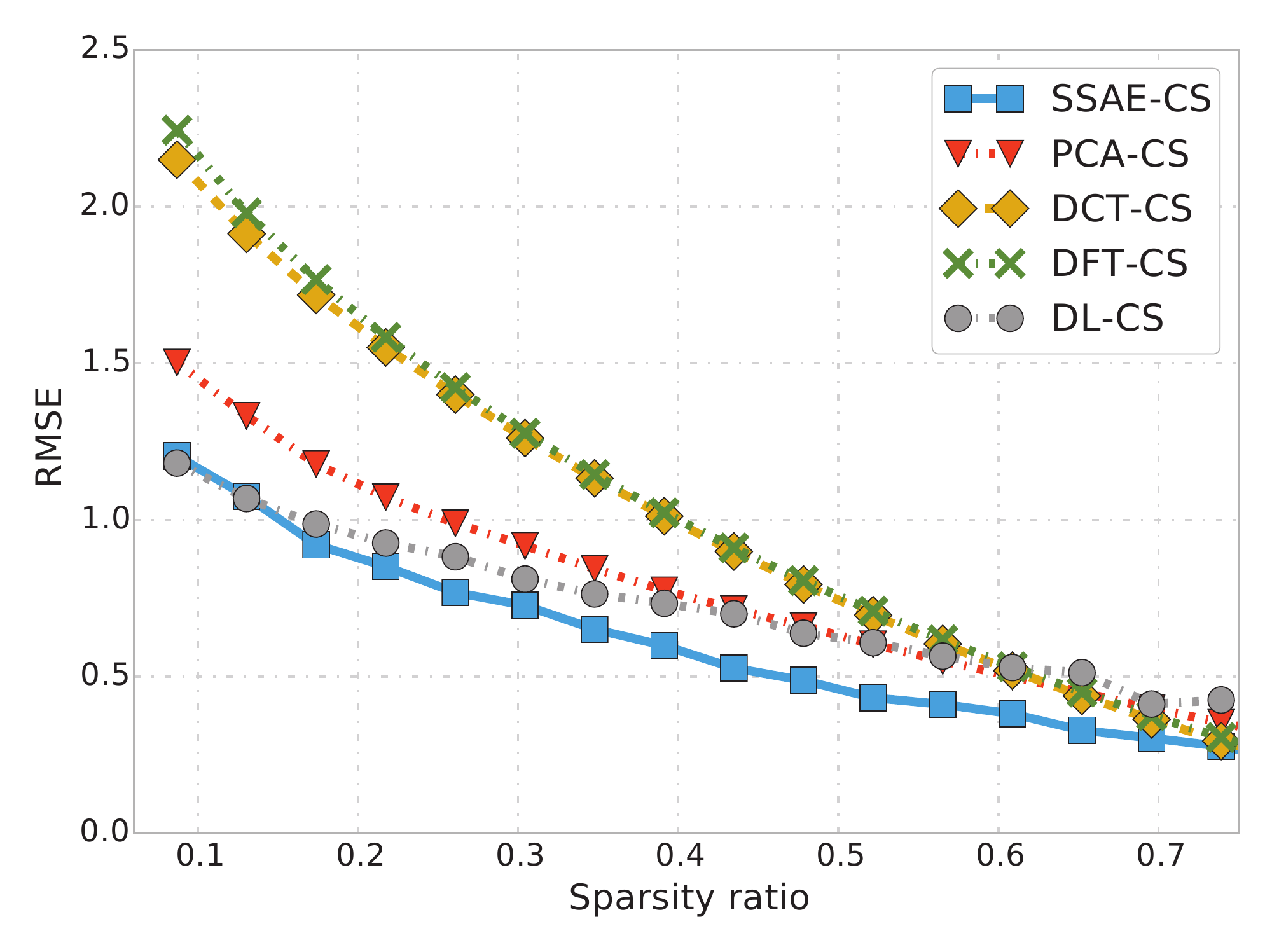}
\par\end{centering}
\caption{\label{fig:mse_sparsity}Root mean square error (RMSE) versus sparsity ratio $\eta=\left(\frac{K}{L}\right)$.}
\end{figure}

Using a difference matrix that captures the difference between adjacent and correlated values as a sparse basis was used in~\cite{quer2009interplay,wu2012situ}. Similar to~\cite{quer2009interplay}, we noted the difference matrix's poor performance in sparsifying the data, and hence it is not included in our comparison analysis.

Figure~\ref{fig:mse_sparsity} shows a comparison between the SSAE recovery performance and other conventional methods including principal component analysis (PCA), discrete Fourier transform (DFT), discrete cosine transform (DCT), and dictionary learning (DL).  These conventional methods are chosen for comparison as they are widely used in the CS literature~\cite{masiero2009data,bajwa2006compressive,luo2009compressive,quer2009interplay,fazel2013random}. Two important observations can be made.
\begin{enumerate}
\item Most sparsity inducing algorithms will achieve a relatively similar recovery error at high values of $\eta$. However, these high sparsity ratio values (e.g., $\eta>0.7$) are not typical in practical applications as the reduction in data size is not noticeable. Therefore, these values cannot be used for CS's applications as the measurement vector size will be similar to the source signal size (i.e., $N\approx M$). On the other hand, SSAE significantly outperforms conventional methods for practical low sparsity ratios and when the nonzero values in the generated sparse codes are required to be minimized.
\item Conventional DL methods (e.g.,~\cite{lee2006efficient,mairal2009online}) use the $\ell_1$ minimization to model the raw data as linear combinations of sparse bases. In this paper, we used the scikit-learn library~\cite{pedregosa2011scikit} for testing the dictionary learning method in which the coordinate descent method is used to find the LASSO problem solution. Similar to our algorithm, the scikit-learn's implementation enables setting the required sparsity ratio by defining the number of nonzero coefficients in the sparse code, while we set the remaining parameters to their default values. We normalize the data to a zero mean and a unit variance before learning the dictionary model. In addition to the slightly better performance, we also noticed that the learning time of the SSAE method is also shorter than the DL method. This is significant for large-scale WSNs.
\end{enumerate}

\subsection{Noisy Data}\label{sub:unreliable_network}

Sensors may report imprecise measurements due to external noise sources, inaccurate sensor calibration, unstable power supply, and imperfect node design~\cite{ni2009sensor}. In this section, we assume that noise values are independent Gaussian variables with zero mean and variance $\sigma_{z}^{2}$ such that $z\sim\mathbb{N}\left(0,\sigma_{z}^{2}I_{N}\right)$, where $\mathbf{z}\in\mathbb{R}^{N}$ is an added noise vector. We noticed that the SSAE method does not only allows the compression of the sensors' data, but it also helps in estimating the noiseless data vector of the physical phenomenon $\mathbf{x}^{*}\in\mathbb{R}^{N}$.

An overcomplete sparse representation is achieved when the number of hidden layer's neurons (sparse code's size) is greater than the input layer's neurons (i.e., $L>N$). However, the measurement vector's size $M$ of CS is proportional to the sparse code's size as in (\ref{eq:cs_reconstruction}). Therefore, the number of nonzero items must be minimized, and less nonzero coefficients are defined in the overcomplete sparse code. On the other hand, using more neurons in the hidden layers can result in the overfitting problem~\cite{liu2008optimized}. Overfitting degrades the neural network's reconstruction performance and increases the learning time of the parameters $\boldsymbol{\theta}\doteq\left[\mathbf{W}^{(1)},\mathbf{W}^{(2)},\mathbf{b}^{(1)},\mathbf{b}^{(2)}\right]$. Table~\ref{tab:overcomplete_network} summarizes the experiments of using overcomplete sparse representation. The results also include the case of adding external noise $z\sim\mathbb{N}\left(0,I_{N}\right)$ to sensors' measurements. This shows that the overcomplete case is useful in unreliable network to reduce the noise effects while producing sparse codes. However, in noise-free networks, using overcomplete codes can degrade the sparsity-inducing algorithm performance due to the overfitting problem.

\begin{table}
\centering{}\caption{\label{tab:overcomplete_network}System performance with different numbers of hidden neurons.}
\begin{tabular}{|>{\centering}m{0.6cm}|>{\centering}m{0.6cm}|>{\centering}m{0.6cm}|>{\centering}m{0.6cm}|>{\centering}m{1.9cm}|>{\centering}m{1.9cm}|}
\hline 
$L$ & $K$ & $\gamma$ & $M$ & RMSE (no external noise) & RMSE (noise $\sigma_{z}^{2}=1$)\tabularnewline
\hline 
\hline 
23 & 5 & 0.2 & 12 & 0.987 & 1.522 (unreliable)\tabularnewline
\hline 
25 & 5 & 0.25 & 12 & \textbf{0.930 (best)} & 1.512 (unreliable)\tabularnewline
\hline 
30 & 4 & 0.5 & 12 & 0.982

(overfitting) & \textbf{1.259 (best)}\tabularnewline
\hline 
32 & 4 & 0.6 & 12 & 1.027 (overfitting) & 1.338\tabularnewline
\hline 
\end{tabular}
\end{table}

%%%%%%%%%%%%%%%%%%%%%%%%%%%%%%%%%%%%%%%%%%%%%%%%%%%%%%%%%%%%%%%%%%%%%%%%%%%%%%%%%%%%%%%%%%%%%%%
\section{Summary}\label{sec:summary}

In this paper, we have introduced a sparsity-inducing algorithm for data aggregation of non-sparse signal in wireless sensor networks. The proposed method consists of three steps: data collection, offline training and modeling, and online sparse code generation. The modeling scheme is based on a neural network with three layers, where the sparse codes are exposed at the hidden layer's neurons. A cost function is introduced as a sparsity nomination scheme. Then, a shrinking mechanism is used to switch off the least dominant neurons in the hidden layer, while asserting on the number of generated nonzero values in the sparse code. The resulting scheme can be used in many applications such as in compressive sensing-based data aggregation schemes.

For future research, we will analytically study the energy consumption and computational burdens of the proposed scheme.

\bibliographystyle{IEEEtran}
\bibliography{sparse_coding}

\end{document}